\documentclass[english]{article}
\usepackage{mathptmx}
\usepackage[T1]{fontenc}
\usepackage[latin9]{inputenc}
\usepackage{geometry}
\usepackage{amsmath}
\usepackage{amssymb}
\usepackage{amsthm}

\usepackage{xcolor}
\usepackage[authoryear]{natbib}

\makeatletter

\usepackage{soul}

\AtBeginDocument{
  
}

\makeatother

\usepackage{babel}
\begin{document}
\title{Counterexamples to "The Blessings of Multiple Causes" by Wang and Blei}
\author {Elizabeth L. Ogburn, Ilya Shpitser, and Eric J. Tchetgen Tchetgen}

\maketitle

\begin{abstract} 
This brief note is meant to complement our previous comment on "The Blessings of Multiple Causes" by Wang and Blei (2019).  We provide a more succinct and transparent explanation of the fact that the deconfounder does not control for multi-cause confounding. The argument given in \cite{wang2019blessings} makes two mistakes: (1) attempting to infer independence conditional on one variable from independence conditional on a different, unrelated variable, and (2) attempting to infer joint independence from pairwise independence.   We give two simple counterexamples to the deconfounder claim. 
\end{abstract}

\textbf{Update (April, 2020):} In a response to the counterexamples below, \cite{wang2020towards} (https://arxiv.org/abs/2003.04948) propose two new assumptions to justify the deconfounder.\footnote{Assumption 1 is stronger than JASA Definition 4, which it is meant to replace.  Definition 4 allows $P(A_j|V_j)$ to be a point mass, while Assumption 1 rules that out.  Because Definition 4 deals separately with each $V_j$, the component of $U$ that affects cause $j$, it does not ensure that the collection of $V_j$'s comprise a multi-cause separator.  This is, again, because pairwise independence is weaker than joint independence, so the "minimal sigma algebras" required for the pairwise independences in (40) may not suffice for joint independence to hold. Because Definition 4 deals separately with ignorability for each cause, it does not ensure that joint ignorability for the collection of causes holds conditional on the collection of $V_j$'s.  

After redefining "substitute confounder" according to the new Assumption 1, JASA Definition 5 and Assumption 2 are equivalent.} 
We agree that these assumptions suffice for a factor model of the causes to control for confounding.  Indeed, the claim in this document is unsurprising and uncontroversial.  It can be summarized as: "If an unmeasured confounder $U$ is distributed according to a probabilistic factor model on the causes, and if that probabilistic factor model is uniquely pinpointed by a deterministic function of the causes as the number of causes goes to infinity, then $U$ can be recovered by a probabilistic factor model on the causes as the number of causes goes to infinity."  This claim is not entirely trivial; the proof requires the additional assumption that $U$ is a multi-cause separator under which WB prove that $U$ is the \emph{unique} solution to a probabilistic factor model on the causes, in the limit as the number of causes goes to infinity.  

However, these new assumptions raise new problems for the Theorems in the original JASA paper \citep{wang2019blessings}.  In particular, weaknesses in the parametric identification strategies that \cite{d2019comment} pointed out become logical impossibilities.  Consider conditional expectations like $E[Y|A=a,Z=z ]$, which appear in the statement of Theorem 6 and the proofs of Theorems 6 and 7 (lines 104 and 109 in the Appendix).  Because $Z=U$ is a deterministic function of $A$, we can write $E[ Y | A=a,  Z=z ] = E[ Y | A = a, f( A ) = z ]$. This is only well-defined only for $z = f( a )$; otherwise the conditioning event $ A = a,   f( A ) = f( a' )$ has probability $0$.  If, as was sometimes taken to be the case in the JASA paper, $Z=f(a)$ is a mere stand-in for $U$, then parametric strategies can be used to overcome the issue of undefined conditioning events.  But under the new, clarified assumptions that is not possible:  as a concrete example, no parametric strategy can make $E[Y|A=1, A^2=10]$ well-defined.  Therefore, under the new assumptions the theorems themselves do not hold even though conditional ignorability does.


\textbf{Original text (January, 2020):} We echo the sentiment, expressed by Wang and Blei (henceforth WB), that it is a privilege to be a part of such a vigorous intellectual discussion. We were constrained by time pressure in writing our original comment \citep{ogburn2019comment}, and in the months since then we have have found a more concise, and we hope transparent, way to explain why the deconfounder fails to control for multi-cause confounding.  
We present counterexamples to Lemmas 1, 2 and 3, along with a simple explanation of the flawed reasoning in these lemmas. (Counterexamples to Lemma 4 follow from, but are more or less obviated by, the fact that Lemmas 1, 2, and 3 do not hold.)  



Suppose we are interested in the effect of a collection of potential causes or treatments on an outcome. The foundational claim of the "The Blessings of Multiple Causes" is that, if we can find a (latent) variable that makes the causes mutually independent, then it suffices to control for all multi-cause confounding.  This variable is the "deconfounder" from which the proposal takes its name.  This claim is incorrect. It relies on two incorrect inferences: (1) attempting to infer independence conditional on one variable from independence conditional on a different, unrelated variable, and (2) attempting to infer joint independence from pairwise independence.

This claim is made formally in Lemmas 1 and 2, and in Lemma 3 which follows almost immediately from Lemmas 1 and 2.  It is described informally in Section 2.2 of WB (2019), e.g. in this passage:

\setlength{\leftskip}{1cm}
\setlength{\rightskip}{1cm}

Here is the punchline. If we find a factor model that captures the population distribution of assigned causes then we have essentially discovered a variable that captures all multiple-cause confounders. The reason is that multiple-cause confounders induce dependence among the assigned causes, regardless of how they connect to the potential outcome function. Modeling their dependence, for which we have observations, provides a way to estimate variables that capture those confounders. This is the blessing of multiple causes.



\setlength{\rightskip}{0pt}
\setlength{\leftskip}{0pt}

But this is incorrect; that the causes are independent \emph{from one another} conditional on $Z$ does not imply that the causes are collectively independent \emph{from a new variable altogether}---a potential outcome---conditional on $Z$.  

In fact, the claim described above is a simple statement about the relationships among conditional and joint probability distributions, and it may clarify matters to state the claim without any reference to potential outcomes or to causality.  To that end, we will use $A_1,A_2,...$ to represent the causes, $W$ to stand in for a potential outcome $Y(a_1,a_2,...)$, and $Z$ to be any variable that renders the causes conditionally mutually independent.  Then the crucial, but incorrect, claim made in Lemmas 1, 2, and 3 of WB (2019) can be stated as follows:

\emph{\textbf{The Deconfounder Claim:}}

\emph{Assume}
\begin{equation}
\begin{split}
   & \text{There exists } U \text{ such that for all } j \footnotemark
 \\
  & (i) \hspace{3bp} A_{j} \perp W | U, \\
  & (ii) \hspace{3bp}  A_j \perp (A_1,...,A_{j-1},A_{j+1},...) | U 
\end{split}
\end{equation}
\footnotetext{There is an additional technical requirement which we will discuss in footnotes throughout in order to simplify the main text: 

\emph{(iii) No proper subset of $\sigma(U)$ satisfies (ii).  }

We interpret this to mean that there is no random variable $X$ measurable with respect to a proper subset of $\sigma(U)$ such that (ii) holds conditional on $X$ instead of $U$.}
\emph{and}
\begin{equation}
(A_{1},A_{2},...) \text{ are mutually independent given } Z.
\end{equation}
\emph{Then}
\begin{equation}
(A_{1},A_{2},...)\perp W|Z.
\end{equation}

Although we have abstracted away from causality and potential outcomes here, when the $A$s are causes and $W=Y(a_1,a_2,...)$ is a potential outcome, (1) and (3) are important statements about confounding and ignorability.  Assumption (1) is a simplified but stronger version of WB's "no single-cause confounding''
assumption (Definition 4 of WB, 2019).\footnote{The version in WB allows this to hold for a different $U_j$ for each $j$. We make this simplified assumption without loss of generality, because it implies WB's version.}  All of our arguments below can be conditioned on any observed single-cause confounders.    Assumption (2) is the assumption that the causes come from a factor model with latent factor $Z$ (see Equation 5 and Definition 2, Equation 35 in WB, 2019).  The conclusion (3) is conditional ignorability or "weak unconfoundedness;" it says that $Z$ suffices to control for any multi-cause confounding of the effect of $(A_1,A_2,...)$ on $Y$.  In the absence of single-cause confounders, as we assume throughout, $Z$ suffices to control for all confounding and to identify causal effects.

But (3) does not follow from (1) and (2).  The Deconfounder Claim is false because it attempts to infer independence conditional on $Z$ from independence conditional on $U$, and because it attempts to infer joint from pairwise independence.
Clearly (3) does not follow from (2); you cannot learn what independences hold in the joint distribution of $(A,W,Z)$ from the marginal distribution of $A$ and $Z$.  And (1) cannot help, because $U$ may be entirely unrelated to $Z$.  Even if we were to stipulate that $U=Z$ in (1), (i) is about pairwise independence which does not imply the joint independence of (3).



In order to highlight these two inferential errors, consider the case with only two causes.  Then the Deconfounder Claim can be restated as follows:

Assume
\begin{equation}
\begin{split}
   & \text{There exists } U \text{ such that} \footnotemark\\
  & (i) \hspace{3bp} A_{1} \perp W | U \text{ and } A_2 \perp W | U, \\
  & (ii) \hspace{3bp}  A_1 \perp A_2 | U\\
\end{split}
\end{equation}
\footnotetext{Plus the technical requirement that (iii) no proper subset of $\sigma(U)$ satisfies (ii).}
and
\begin{equation}
A_{1}\perp A_{2}|Z.
\end{equation}
Then
\begin{equation}
(A_{1},A_{2})\perp W|Z.
\end{equation}

But (6) does not follow from (4) and (5).  Because (4)(i) does not condition on $Z$, it does not rule out
the case that, e.g., $A_1\not\perp W|Z$, in which case the conclusion does not hold. 
Even if we replaced $U$ with $Z$ in (4) joint independence does not follow from pairwise independence statements such as (4) and (5).  This can be seen from the counterexample below, which is adapted from the canonical example of pairwise without joint independence:

\textbf{Counterexample 1 }

$Z\sim Ber(0.5)$, $U$ is null (e.g. a constant).

$A_{1},A_{2},$ and $W$ are independent of $Z$, with $(A_{1},A_{2},W)=\left\{ {\begin{matrix}(0,0,0) & {\text{with probability}}\ 1/4,\\
(0,1,1) & {\text{with probability}}\ 1/4,\\
(1,0,1) & {\text{with probability}}\ 1/4,\\
(1,1,0) & {\text{with probability}}\ 1/4.
\end{matrix}}\right.$

Then
\begin{enumerate}
\item (i) $A_{1}\perp W,A_{2}\perp W$ and (ii) $A_1 \perp A_2$,\footnote{$\sigma(U)$ is trivial and has no proper subsets, so (iii) holds trivially.}
\item $A_{1}\perp A_{2}|Z$
\end{enumerate}
but $(A_{1},A_{2})\not\perp W|Z$. 
$\blacksquare$

A peculiarity of this counterexample is that (2) holds even without conditioning on $Z$; in this second example $Z$ is required to make $A_{1}$ and $A_{2}$ independent.  We let $U=Z$ to show that the deconfounder claim would fail to hold even if assumption (1) were modified to relate to $Z$.

\textbf{Counterexample 2}

$U=Z\sim Ber(0.5)$

When $Z=0$, $(A_{1},A_{2},W)=\left\{ {\begin{matrix}(0,0,0) & {\text{with probability}}\ 1/4,\\
(0,1,1) & {\text{with probability}}\ 1/4,\\
(1,0,1) & {\text{with probability}}\ 1/4,\\
(1,1,0) & {\text{with probability}}\ 1/4.
\end{matrix}}\right.$

When $Z=1$, $(A_{1},A_{2},W)=\left\{ {\begin{matrix}(0,0,0) & {\text{with probability}}\ 1/4,\\
(0,2,1) & {\text{with probability}}\ 1/4,\\
(2,0,1) & {\text{with probability}}\ 1/4,\\
(2,2,0) & {\text{with probability}}\ 1/4.
\end{matrix}}\right.$

Then
\begin{enumerate}
\item (i) $A_{1}\perp W|Z,A_{2}\perp W|Z$ and  (ii) $A_1 \perp A_2 | Z$ \footnote{(iii) holds trivially because $Z$ is the only conditioning event that can make (ii) hold.} 
\item $A_{1}\perp A_{2}|Z$
\end{enumerate}
but $(A_{1},A_{2})$ $\not\perp$$W|Z$.
$\blacksquare$

Requiring $Z$ to be a deterministic function of $A_{1},A_{2},...$, as is the case in WB's factor model, does not solve these problems. Consider a toy example with 3 causes, and let $Z=f(A_{1},A_{2},A_{3})=A_{3}$. The counterexamples above still hold, with $Z=A_3$.
(This obviously has overlap issues, but it generalizes easily to a
setting, akin to Theorem 7, where $Z$ is a function of $A_{3},A_{4},...$
and we're interested in the joint effect of $A_{1}$ and $A_{2}$. See Remark 3.)

\textbf{Remark 1.} \emph{Methods like the deconfounder work in some settings, but rendering the causes conditionally independent is a consequence, not a driver, of their success.}
Although the deconfounder claim is incorrect as stated above and in WB (2019), e.g. in Section 2.2 and in Lemmas 1 and 2, we do not dispute the fact that factor models can control for unmeasured confounding under additional parametric assumptions.  Even in these settings, it's not the case that $Z$ controls for confounding \emph{because} it renders the causes conditionally independent.  In most of the settings in which methodology like the deconfounder works, unmeasured confounding is due to clustering or some other structure (e.g. kinship structure in GWAS) that is common to each random variable in the collection $A_1,A_2,...,Y(a_1,a_2,...)$. A factor model may control for confounding because it captures this common structure.  A (sometimes unacknowledged) assumption of these methods is that the only source of dependence among the causes is common structure, but if there were a variable that rendered the causes conditionally independent yet failed to capture the common structure, that variable would not suffice to control for confounding.   

\textbf{Remark 2.} \emph{Probabilistic factor models.} In the examples above, there is a trivial factor model for the causes $A$, as in Definition 2 of WB (2019).  

\textbf{Remark 3.} \emph{Overlap.} 
Many of the results in WB (2019) require an \emph{overlap} assumption (Equation 6 of WB, 2019).  This assumption is standard for causal inference and is often called \emph{positivity}.  Overlap is entirely separate from ignorability/unconfoundedness, and is irrelevant to Lemmas 1 through 3, and also to Lemma 4.  Overlap is not referenced or used in the statements or proofs of any of these lemmas. That overlap is orthogonal to Lemmas 1 through 4, and for unconfoundedness specifically, is evident in the statement of Theorem 7, which does not require overlap but does hinge on Lemma 3 (and therefore Lemmas 1 and 2) and Lemma 4.  Therefore, when discussing whether unconfoundedness follows from the construction of the deconfounder, we are free to ignore overlap.  
That said, it is easy to modify Example 1 to meet the overlap assumption, simply by introducing additional causes and defining $Z$ to be one of WB's allowable factor models on $(A_3,A_4,A_5,...)$.

\textbf{Remark 4.} \emph{Theorems 6 and 7.} Theorem 7 relies on the Deconfounder Claim for identification, and since that claim is incorrect, so is Theorem 7.  But Theorem 6 (and possibly also Theorem 8, which we did not analyze) is largely correct despite the fact that the proof in WB (2019) is incorrect (it relies on Lemmas 3 and 4, which are false).  This theorem holds because of the strong parametric assumptions that the confounding
variable is a clustering indicator and that the treatment effects are constant across clusters (no treatment-confounder interaction).  This is related to the literature on using mixed effects models to control for unmeasured cluster-level confounding \citep{brumback2010adjusting, brumback2017use, goetgeluk2008conditional, imai2019should}. 

\section*{Acknowledgements}
We would like to thank Alex D'Amour, Jeff Leek, Wang Miao, and Alex Volfovsky for helpful discussions.

\bibliographystyle{jasa}
\bibliography{refs}

\end{document}